# A Decentralised Real Estate Transfer Verification Based on Self-Sovereign Identity and Smart Contracts

Abubakar-Sadiq Shehu[1,2] [a], António Pinto[3] [b] and Manuel E. Correia[1,3] [c]
[1]*Department of Computer Science, Faculty of Science, University of Porto, Porto, Portugal.*
[2]*Department of Information Technology, FCSIT, Bayero University Kano, Kano, Nigeria.*
[3]*CRACS & INESC TEC, Porto, Portugal.*

Abstract: Since its first introduction in late 90s, the use of marketplaces has continued to grow, today virtually everything from physical assets to services can be purchased on digital marketplaces, real estate is not an exception. Some marketplaces allow acclaimed asset owners to advertise their products, to which the services gets commission/percentage from proceeds of sale/lease. Despite the success recorded in the use of the marketplaces, they are not without limitations which include identity and property fraud, impersonation and the use of centralised technology with trusted parties that are prone to single point of failures (SPOF). Being one of the most valuable assets, real estate has been a target for marketplace fraud as impersonators take pictures of properties they do not own, upload them on marketplace with promising prices that lures innocent or naive buyers. This paper addresses these issues by proposing a self sovereign identity (SSI) and smart contract based framework for identity verification and verified transaction management on secure digital marketplaces. First, the use of SSI technology enable methods for acquiring verified credential (VC) that are verifiable on a decentralised blockchain registry to identify both real estate owner(s) and real estate property. Second, the smart contracts are used to negotiate the secure transfer of real estate property deeds on the marketplace. To assess the viability of our proposal we define an application scenario and compare our work with other approaches.

## 1 INTRODUCTION

Web 2.0 is one of the many advancements that the Internet has witnessed since its creation. It has seen significant growth, primarily in marketplace and electronic transactions where practically anything, including real estate, can be offered for rent, lease, or sale on marketplaces and blogs (Facebook, Airbnb, Uniplaces and others) for residential, commercial or industrial usage. Globally, real estates are considered one of the most valuable assets (commonly used as collateral for obtaining loans from both formal financial institution such as banks and informal credit providers) (Yadav and Kushwaha, 2021), contributing a significant share to Governments GDP. For example in 2020 it accounted for about 7.5% of the Chinese economy (Pain and Rusticelli, 2022) and 17.5% in the USA for fixed investment and total housing spending in 2019, real estate commercial properties contributed 3.1% (EUR 452 billion) to the European economy, a value comparable to the combined contribution of the automotive and telecommunications industries (EPRA, 2020). More so, real estate is the primary component of agriculture and, as a result, it is inextricably related to food security. While properties are easily available for buyers choice, determining the authenticity and correctness in real estate marketplaces can be a daunting challenge. Typically, interactions with individuals in real estate transaction requires proper identification, often with a large amount of paperwork issued by different institutions, which takes a lot of time to confirm from the issuers. Additionally, these services are centralised and siloed, therefore, any failure on the process can completely undermine the progress of an entire transaction.

To curb the aforementioned issues, Governments, businesses and researchers have been working on improving the security of digital commercial transactions and associated treatments of personal data. This led to the idea of Web 3.0 (Ragnedda and Destefa- nis, 2019) which introduces fairer and more secure communication and data exchange methods, enabling users to be sovereign on their own identity, through decentralisation and the use of blockchains.

[a] https://orcid.org/0000-0002-2894-6434
[b] https://orcid.org/0000-0002-5583-5772
[c] https://orcid.org/0000-0002-2348-8075

Self-sovereign identity (SSI), smart contract and distributed storage are among the recent advances in secure data exchange and verification. SSI provides a method for data owner sovereignty, in most cases it uses blockchain for secure storage and verification of transactions, while smart contract provides an ecosystem for secure transaction negotiation, both do not need third party verification. This paper proposes an SSI, smart contract and distributed storage based real estate verification framework, whose aim is to prevent fraud in transfer processes, ensure secured identity verification and mitigate tax evasion in transactions.

**Contribution** A good number of research works some of which are discussed in Section 2, aimed to address these issues have focused on decentralising the real estate registration process, while some other works focused on identity, fraud and verification issues. Therefore, we present a framework that ensures the secure creation, verification of VC and a trans- fer process that manages real estate transaction based on smart contracts (szabo and Nick, 2018).This works was inspired by our previous works (Shehu et al., 2020)(Shehu et al., 2019) and is guided by GDPR principles in (Voigt and Von dem Bussche, 2017). To our knowledge no work has combined SSI, smart contracts and IPFS peer-to-peer distributed file system (Benet, 2014), to address real estate verification, property transfer and associated secure data storage issues.

The contributions of this paper are as follows:

1. **Review Study:** We did an extensive survey of related state of the art works on dematerialised real estate management that uses blockchain and other SSI components. We identified some limitations that we address on the framework for securing real estate transactions that we are proposing in this paper.
2. **System Framework**: We propose an SSI and smart contract framework for verifying real estate transfer processes. The SSI layer of the framework is used to define methods and the process of acquiring VCs (electronic equivalence of our physical documents/credentials e.g national cards and others) (Consortium et al., 2019) for actors (real estate, owner, buyer and marketplace) that are linked and verifiable on a distributed ledger. While smart contracts are used to implement the negotiation, the real estate owner remains in control and can deploy those smart contracts at will on a marketplace, for intending buyers to engage with. We enhance our framework with the IPFS, a secure distributed storage system that we use to store a hash and an encrypted copy of the complete real estate description.
3. **Demonstration pilot**: To conceptualise the proposed framework, we present an application scenario for verification of real estate transfer, and analyse the characteristics of each component.
4. **Evaluation:** From a functional point of view we elaborated the strengths and possible weaknesses of our proposal and propose future work.

**Outline** This paper is structured as follows: Section 3 provides a background study and associated technologies. We discuss related works of the proposed framework in Section 2. In Section 4 we provide an overview of our proposed solution and discuss it components and actors. Section 5 discusses application scenario of the proposed framework for use of SSI and smart contract transfer method. We discuss security characteristics and robustness of the proposed work and the three main components in Section 6. We conclude the paper in Section 7 and discuss future work.

## 2 Related works

While emphasising the need for adopting blockchain solution in real estate to enable and facilitate low-cost P2P commercial interactions, the authors in (Norta et al., 2018) proposed a B2B crowdfunding platform for commercial real estate leveraging the Evareium digital real estate fund token system, with quality goals such as overall system security, seamless information flow between platform sub-infrastructures. In a bid to address fraud in buying, selling and tempering with real estate records, and eliminate the use of centralised land registry method in Saudi Arabia, the authors in (Ali et al., 2020) proposed a blockchain based framework using a permissioned hyperledger fabric blockchain, in which users can utilize smart contracts to buy, sell, and update property records. The system keeps track of all prior purchasers and sellers of a property, which can be validated on the network. The work in (Gupta et al., 2020) proposed a real estate investment solution introducing liquid- ity using blockchain and a tokenised special purpose vehicle for investors to purchase ERC 777 standard security tokens at their leisure. Drawbacks of the method include complete reliance on third-parties for identity verification of investors, likewise the paper work can be time consuming and prone to forgery. The authors in (Kothari et al., 2020) proposed a transparency tamper proof platform for real estate. The system is composed of an owner, buyer and validator.

A major drawback of this system is the over reliance on third party and authenticity of advertised properties are not verified. To prevent fraud in traditional real estate transactions, the work in (Yadav and Kushwaha, 2021) proposed a blockchain-based system for digitising real estate transactions to reduce the risks of fabricating documents and other fraudulent behaviors. In their claim, the work employs a consensus mechanism that minimises multi casting node overhead transmissions by 50 %. The work in (Mendi et al., 2020) proposed a blockchain framework to reduce the tax evasion in real estate transactions within Turkey, using framework used hyperledger and incorporating all parties involved in the transfer of ownership; land registry office to monitor the sale, bank for payment of funds, municipality for tax on buyer and seller. The authors in (Bhanushali et al., 2020) proposed a system to address lost or damaged deeds of real estate using a smart contract. To sell or buy in real estate, a user fulfills the smart contract's requirements and receive a digital deed, which is then uploaded as a new block in a blockchain. A major drawback of this method is the non verification of the property with a users identity. Any user who is able to present proof of a real estate fulfills the smart contract requirement.

## 3 Preliminaries

**Self Sovereign Identity:** SSI and smart contract are among the advances of Web 3.0 and data decentralisation. SSI provides a decentralised data structure, where data owner can be self-reliant from services, free to create their VC (digital equivalence of physical documents), revoke or delegate them at will. Decentralised identifiers, VCs, and immutable registers (blockchain) have been major drivers of SSI. In SSI, a blockchain registry is a distributed ledger that is globally accepted by SSI actors as a source of truth, where they can easily confirm the validity and authenticity of a credential without revealing its content. Since its introduction, SSI methods have been implemented in a couple of areas to address trust, privacy, and security issues in IoTs, ehealth, finance, real estate among others. With SSI, users are able to limit the practice of data extraction and personal information collection by services without their consent.

The SSI leverages on a trust triangle consisting of three actors (issuer, holder and verifier). Issuer is an entity that is able to verify identity attributes of a holder and issue a VC. A holder is an identifiable entity that receives credentials from issuers, then presents them as proof of claim to a verifier. Veri-fiers are service provider (SP), that sets access policy for holders and provide their services based on the policies. The relationship between these actors is depicted in **Figure** 1.

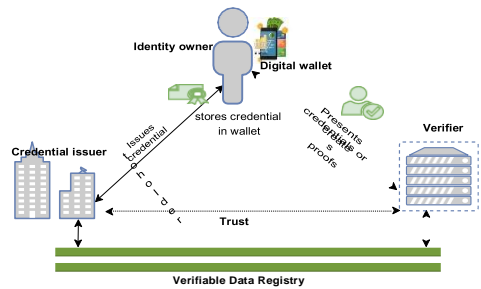

Figure 1: Self-Sovereign identity trust triangle.

**Blockchain:** In 2009, blockchain attracted the world's attention through the first known real-world application of bitcoin cryptocurrency (Nakamoto, 2008), as a distributed and decentralised system consisting of an immutable public database, that are cryptographically hashed in peer-to-peer public transactions. A block is made up of a series of transactions that are not managed by a single centralised organisation, but are instead publicly available and trustworthy to all network users. Traditionally, blockchains are classified either as public and private (Soltani et al., 2021), or as permissioned and permissionless.

Blockchain shares a consensus algorithm that allows immutable transactions to be completed and synchronised, it also generates an ordered list of stored and associated information through a chain of blocks that usually contain the previous hash block, data content, participant signature and timestamp. The previous hash block causes the information in the blockchain to remain immutable.

**Smart Contract:** The first published work on smart contract was by Nick Szabo in 1994 (szabo and Nick, 2018). Smart contracts are digital contracts that are stored on blockchain they are akin to physical contracts. They inherit the properties of blockchain such as immutability, distributed and trustless network. Typically smart contracts are used to automate the execution of an agreement that is triggered through protocols and conditions, so that all participants are immediately certain of the outcome, without the involvement of an intermediary or the loss of time.

**Interplanetary File System (IPFS):** IPFS is an open-source set of protocols (Benet, 2014) which integrates various existing concepts including peer-to-peer (P2P) networking, Linked Data, and other do-mains to allow participants share file fragments. Content on IPFS is uniquely labeled and addressed using a multihash, which is a self-describing datatype that incorporates features from git's versioning model, cryptographic hashing, and Merkle Trees to make file retrieval easier. Contents are identified and accessible using names rather than location.

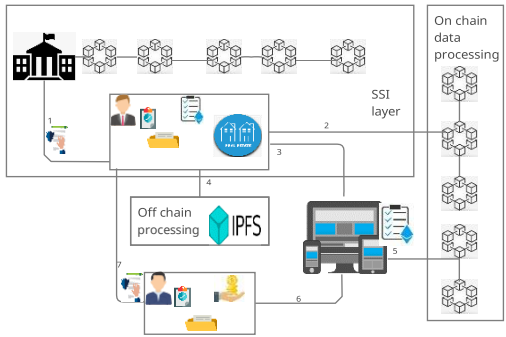
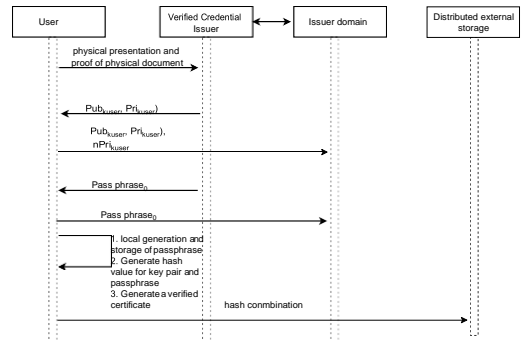

Figure 2: A high-level architecture of proposed system

Figure 3: Owner registration process

# 4 A solution for verified real estate transaction based on SSI and smart contracts

In this section we analyse the our proposal as follows: firstly, the overall architecture is described with all actors, then, identity management and verification process is analysed, lastly, the ownership transfer process and data storage procedures are discussed. The generic overview of the proposed framework is depicted in **Figure** 2. It is divided into three nodes; identity management node, on-chain data processing and off-chain external data storage. The identity management node which follows an SSI method.

On chain data processing node consist of two Ethereum based smart contracts for negotiating the transfer of real estate. While the off chain external storage layer stores a comprehensive digital information of the real estate (description, pictures, videos) and ownership transfer history.

## 4.1 Identity management process

Securely verifying the real estate and users (market place, owner and buyer) in a decentralised manner forms the core of this work.

**User Identification:** Considering the fact that the asset at hand is a physical property, we initiate the identification process with a pre-requisite physical identification of the owner as well as the real estate. In the course of this work we assume the credential issuer to be a Government/state office which owns a distributed and verifiable data registry that stores digital information of residents and businesses. At the state office, a user presents physical documents containing personal attributes (e.g name, date of birth, address and others) for identification on which due diligence is conducted. For subsequent processes, an owner may initiate a registration on the registry's mobile or web domain by creating a user name and private key. Either way two postal mails are sent to the user's registered address, first mail contains a public and temporary private key that has to be changed. A user would then have to access the Government's domain to register the credentials received and change associated key. Once done, the system automatically triggers the request for the second mail which contains a unique and changeable mnemonic passphrase. Owner's account is still under verification until the passphrase is received and changed. Once completed owner is able to request for physical and equivalent VC. The VC contains a public key ($Pub_{kowner}$) that is locally stored on owner's digital wallet, a private key ($Pri_{kowner}$) that unlocks owner's account and a hash $hf$ of the key pair that serves as the identity of the owner VC$_{owner}$=( $hf$ ($Pub_{owner}$, $Pri_{owner}$), $Pri_{owner}$, $Pub_{owner}$). The hash value is deposited in decen- tralised the registry of the issuer to uniquely identify owner. On the decentralised state mobile or Web do- main, owner logs-in with their self generated private key and confirms the already provided attributes dur- ing the registration. From the key pair, owner is able to derive new hash credentials for unique identifica- tion that will be case specific, and even choose from an array of services (public/private) provided, with which they want to link their identity to. This process is described shown in **Figure** 3.

**Real estate identification:** Been a landed property, real estates are owned by private or state entities, therefore, the identification process follows similar approach of owner identification. We assume a real estate is newly acquired, therefore the owner needs to obtain a VC for it, and bind it to their own VC. At the registry office, the owner presents their VC with proof of ownership (such as deed document or certificate of ownership) that fully describes the prop- erty, the map location, purchase receipt with preced- ing owners identities (VC hashes) and a plan for the property if any. The land registry office dispatches two address verification mail; first to the preceding owner of the property and second to the verified ad- dress of the owner, which contains a key pair gener- ated from the owner's public key and a changeable password for access to land registry domain to com- plete the registration. Once received, the preceding owner has to confirm the transfer of property using owned VC. Thereafter, the owner is able to link the property's public key to VC and locate it on the land registry map (description of this property is greyed out and saved to a distributed external storage). A hash is generated from the key pair, property loca- tion on the map, owner's VC and property public key which is then pinned to external distributed storage that is publicly accessible as shown in **Figure** 4.

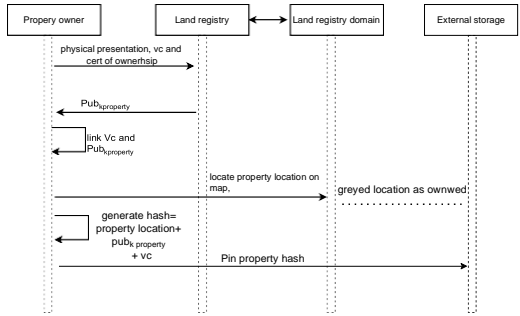

Figure 4: Real estate registration process

## 4.2 On chain data processing

We assume that a buyer/client and decentralised marketplace acquire their VCs using same process with real estate owner. The proposed infrastruc- ture is composed of two smart contracts that form the bases for real estate transfer; which are capabil- ity smart contract $SC_{capability}$ and transfer smart con- tract $SC_{transfer}$. Both contracts are fashioned on an Ethereum blockchain.

---
**Algorithm 1** $SC_{capability}$ creation
---
1: **procedure** Owner verification and contract creation process
2:   Input: $acct, AP, B, start, end, minV\alpha\alpha e, DR, R', VC_{property}, IPFS_{hash}, \alpha B$
     Read values
3:   **if** $d, R', B, VC_{property} = o\alpha ner$ **then**
4:     start transfer
5:     **if** $\alpha fet\alpha me \geq 1$ **then**
6:       Create instance of $SC_{capability}$
7:     **else if** $\alpha fet\alpha me \leq 0$ **then**
8:       Reject creation process
9:     **else**
10:      Create $acct, AP, \alpha\alpha\alpha e, DR, st\alpha t\alpha s$
11:    **while** $St\alpha t\alpha s = \alpha ct\alpha\alpha e$ **do**
12:      $Ret\alpha rn \leftarrow IPFS_{hash}$
13:      $Ret\alpha rn \leftarrow SC_{capability}$
---

$SC_{capability}$: This contract is created and owned by the real estate owner, it hold's all necessary informa- tion that identifies them. To determine method for proof of identity, we define $SC_{capability}$ properties in relation to it's attributes and enumerate what action (access request, delegation, transfer of property and revocation) the owner can perform with it. To achieve this we define the access control (AC) state and iden- tify its variables; AC (A, P, R), where A, is a set of public keys ($Pub_{ku}...Pub_{ku..n}$), R (set of all attributes on VC), P (set of all possible permissions). From these variables an owner is able to derive capability properties for their smart contract: ($AP, R', Pp', DR, B$), where $AP \subseteq A \times P$ is access policy, $R' \subseteq R$ is a set of extra context awareness attributes, $Pp' \subseteq P \times p'$ is access scope, $DR$ is the delegation relation and B is set of binary relations between entities (which could be real estate in question or other entities).

$SC_{transfer}$: This contract is deployed by verified market place. Since the market place is providing the platform for the transfer of real estate property, an initial authentication and authorisation process is required.

**Smart contract creation:** To create a $SC_{transfer}$, we assume a marketplace defines access requirements through scope and policies, which are achieved by identity verification processes (identification, authen- tication and authorisation). To initiate these pro- cesses, an owner contacts the market place and re- quest access to $SC_{transfer}$ services by sending a request that includes $Req_{access} \rightarrow SC_{transfer}$: *acct*, *hashId* and $Pub_{Kowner}$ where *acct* is the type of contract the owner wants to deploy (sale, lease agreements, or power of attorney contract), *hashId*, is the Id of the owner who wants to deploy a contract, and $Pub_{Kowner}$ is the public key. We assume that each call to a market place get assigned a dedicated function in the smart contract which serves as the controller who is able to lookup owner's Id on the distributed storage. To further verify the authenticity of the acclaimed property, owner's call is responded to with a challenge policy and requests for a VC linking the owner to the property. Once received, owner looks up on their digital wallet to confirm if verified attributes are sufficient enough for authentication and covers request policies, if yes, owner is required to confirm the release of verified attributes otherwise the process is discontinued with an error message that requires owner to provide more attributes that are not available. Using a digital wallet, user manually wraps $R'$ and $B$, both properties are signed with owner's private key so that marketplace controller may decode it with the shared public key: Owner marketplace: $Pri_{Kowner}(R', B)$. We assume that this exchange takes place over a secured end-to-end communication. Upon receipt, marketplace controller confirms identity attributes on the ledger, signed claim, device and location binding to the owner that made the initial request, that nonce is meant for the specific transaction and timestamp is valid. Once this is confirmed real estate owner is able to create $SC_{capability}$ and an IPFS link as shown in **algorithm 1**, while a controller reviews the contract on marketplace's domain and deploys it as $SC_{transfer}$, this is shown in **Figure** 5.

```solidity
pragma solidity >=0.5.0 <0.9.0;

contract transfer{
    capability[] public capabilitys;
    function createcapability() public{
        capability newcapability = new capability(payable(msg.sender));
        capabilitys.push(newcapability);
    }
}
```

Figure 5: Transfer contract

## 4.3 Off chain data processing

Despite benefits of blockchain, its not without some limitations which include high cost of transaction among others(Hughes et al., 2019), for this reason decentralised storage solutions such as IPFS, SWAMP, BitTorrent, Filecoin among others are used to store huge data that can be expensive to store on chain in conjunction with blockchain system to create a decentralised database. Although ownership of real estate can be transferred or structure modified, it remains till eternity (atleast the land). The immutability property of IPFS serves well for real estate data storage.

During the contract creation process, a real estate owner deposits complete identity information/description of the property off chain on an IPFS network, but shares a hash value of the storage on the contract. A complete picture of the property is provided with street view, and map location. To prevent unwanted re-use of acquired picture, each document downloaded is watermarked with owner's name, requestors Id and market place ID.

## 5 Application Scenario

Leveraging on data decentralisation and trust less network, we foresee that the proposed framework can be applied for the transfer of dematerialised property that posses a VC. However, for clarity we conceptualise our proposal using a real estate transfer contract. We assume that a buyer wants to acquire a real estate property through a smart contract that is hosted by a market place. The process involves the participation of four entities; a real estate owner that deploys $SC_{capability}$ by calling the function of a market place, market place deploys owner's contract using $SC_{transfer}$ on decentralised owned platform, client/buyer a bidder that expresses interest on advertised property and external IPFS storage. As described in **Figure** 6, the components describe the transfer through initialisation, registration, offer and finalisation.

**Initialisation:** This signifies the commencement of a transfer process where a real estate owner cumulate already created $SC_{capability}$ as shown in **Algorithm** 1, and deposits an IPFS link of the property on an IPFS network. A function for $SC_{transfer}$ is then created to deploy the $SC_{capability}$.

**Registration:** As described in **Algorithm** 1, to deploy the functions of $SC_{capability}$, the smart contract checks five conditions which includes; (1) is the contract deployed by the owner (2) is the property linked to the owner on distributed storage (3) is the start time greater than current time (4) the market place also calculates commission and charges on the transaction with government tax and requests for owner's approval before it is finally published (5) the owner is not involved making an offer for transfer process.

**Offers:** Once all conditions are satisfied and the marketplace deploys transfer contract, it becomes visible to the public. An owner also receives a share ID of the contract which gives them upmost control over the contract. or make an offer but has upmost control to start, stop or suspend the process at will. Once the contract is running, a user who has acquired VC through the identification process described in **Section** 4.1, uses their verified certificate to request participation by making an offer (spending an ether) that is equal or higher than current value of the property. Once the offer is made, a user receives a user-specific access to a full description of the real estate property such as videos, picture, maintenance mode, setup and every other information that might be deemed useful for the transfer process. Generally, buyers are able to view current offers of other buyers, likewise the market place. However, only the owner is able to confirm the full identity of a buyer.

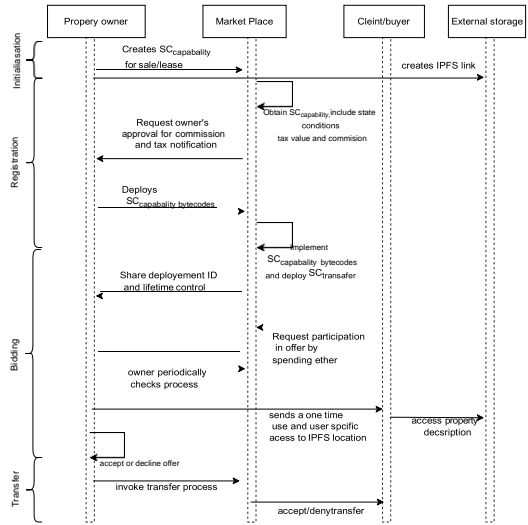

Figure 6: Real estate transfer process

**Transfer:** The owner monitors the transfer process until the lifetime elapses (it can also increase the duration or terminate the entire process when there is a suspicion of fraud). A decision is then made to either accept or reject offers. Buyers who wish to opt out of the transfer process, before the last minutes are fined from the cost paid for the transaction. Once an offer is accepted, a buyer is informed of the process and is requested to proceed with payment before further transfer process is made. To further protect other buyers a percentage of their offer is used as a gas fee for the transaction. However, once a winner is chosen, buyers are requested to proof their identity once again in other to accept a refund for their bids.

# 6 Analysis

From a functional point of view, we analyse the strengths and weaknesses of the proposed framework. We also assess their level of robustness in processing personal information.

**Robustness of Proposed framework:** The proposed SSI and smart contract based real estate verification framework follows a complete decentralisation method. The identification, registration, verification and identity proofing is free from central storage of data. Therefore, there is no means for SPOF.

Issuance, identity verification process and use of VC prevent a malicious attacker from getting hold-on physical documents to claim ownership or steal property. More so, with a trust-less framework, collusion is made difficult as verifier need not to be aware of the physical location of an issuer but can verify the signature of any document it has issued.

**Robustness of Identification and Verification:** These methods are used to capture full attributes of users. Before a real estate VC is issued, user's VC need to have been acquired, therefore linking these documents prevents false claim and other fraud as user proofs ownership of both documents before any transaction.

From verified attributes of a VC, user can generate self acclaimed and context based attributes that are only linked to the VC and verifiable on the distributed register. Therefore, to prevent profiling a user can generate as much context based credentials for transactions without been traced.

**Robustness of Data Processing:** Data decentralisation is among the core parts of the proposed work. Although, one of the drawbacks of blockchain is the use of varying standards which causes fragmentation in adoption, hence interoperability between blockchains only occurs among technologies with same open source implementation or using a trusted party (Bellavista et al., 2021).

To ensure that data processed for the transfer is available, we used the IPFS open source storage which is independently available on Web desktop and provides sufficient memory space that lasted for upto 48hrs. Also, we assume that a completed document signifying the transfer of property is processed off-chain and deposited on the IPFS network.

# 7 Conclusion

In this paper we presented an SSI and smart contract-based real estate transfer framework that can be used to verify the identity of a genuine real estate owner, real estate property and securely control the entire transfer process within a digital marketplace. Our solution uses a decentralised and distributed approach to achieve the seamless verification process with data integrity and personal data protection. The SSI methods ensure that only a verified real estate owner is able to claim ownership of a property using VC that are acquired from an identity verification process and can be validated on a distributed ledger. The smart contract is used to create the transfer process by the owner and is deployed on a decentralised marketplace. A distributed storage is used to store a full description of the property which is accessible using a content identifier. We have also defined an application scenario to test the validity of our proposal. In addition we also provide a robustness and security analysis of our framework with respect to some specific previously identified threats. In the future we plan to fully implement the remaining aspects of the framework, including the issuance of VC, smart contract management, as well as integration of policy makers to the framework for the issuance of tax returns . We also hope to be able to mitigate the non intended use of sensitive documents obtained via the IPFS network.

**ACKNOWLEDGEMENTS** This work is supported by National Funds through the Portuguese funding agency, FCT - Fundação para a Ciência e a Tecnologia, within project LA/P/0063/2020.